\documentclass[preprint,showpacs,preprintnumbers,amsmath,amssymb,a4paper,superscriptaddress]{revtex4}
\usepackage{amsmath}
\usepackage{float}
\usepackage{graphicx, subfigure}
\usepackage{dcolumn}
\usepackage{bm}
\usepackage[font=small,labelfont=bf,labelsep=space]{caption}
\usepackage{refstyle}

\def\nn{\nonumber}

\begin{document}
\title{Multichannel Scattering and Loss Processes of Ultracold Atoms in Anisotropic Harmonic Waveguides}
\date{\today}
\author{Sara Shadmehri}
\email[]{s.shadmehri@iasbs.ac.ir}
\affiliation{Optics and Photonics Research Center, Department of Physics, Institute for Advanced Studies in Basic Sciences
(IASBS),
Gava Zang,
Zanjan 45137-66731,
Iran
}%
\author{Shahpoor Saeidian}
\email[]{saeidian@iasbs.ac.ir}
\affiliation{Optics and Photonics Research Center, Department of Physics, Institute for Advanced Studies in Basic Sciences
(IASBS),
Gava Zang,
Zanjan 45137-66731,
Iran
}%
\author{Vladimir S. Melezhik}
\email[]{melezhik@theor.jinr.ru} \affiliation{Bogoliubov
Laboratory of Theoretical Physics, Joint Institute for Nuclear
Research, Dubna, Moscow Region 141980, Russian Federation}%

\date{\today}

\begin{abstract}\label{txt:abstract}
We have developed the general grid method for multi-channel
scattering of bosonic atoms inside a harmonic waveguide with
transverse  anisotropy. This approach is employed to analyze
elastic as well as inelastic multi-channel confined scattering.
For the elastic scattering, the effects of the range and form of
interatomic potential and the waveguide anisotropy on the
confinement induced resonance are studied. We have also
investigated quantitatively the reactive rate constant in confined
atom-atom collisions. It is found that a slight
anisotropy to the confining trap  considerably enhances the
reactive rate constant in multi-channel regime.
\end{abstract}

\maketitle

\section{Introduction}
Ultracold low-dimensional atomic gases have been intensively
studied over the past two decades
\cite{Kinoshita,Paredes,Haller2009, Chin2010, Kohler2006}. Experimentally, they
are fabricated by employing optical traps which can confine the
system in one, two or three dimensions \cite{Bloch2008}. These low
dimensional systems now provide an ideal experimental tool to test
fundamental problems of few- and many-body physics. The atomic
interactions can be tuned in such structures by varying the trap width or with the help of magnetic Feshbach
resonance by varying the external magnetic field and thereby one
can reach the so-called confinement induced resonance (CIR) which
was first predicted by Olshanii \cite{Olshanii}.
\\
CIRs were confirmed in several experiments \cite{Kinoshita,
Paredes, Haller2009, Gunter2005, Haller2010}. Particularly, in a
prominent experiment \cite{Haller2010}, CIR properties were studied
for an ultracold quantum gas of Cs atoms confined in a quasi-1D
geometry with transverse anisotropy. Observed splitting of the
peek in atomic losses near the CIR, which repeated the splitting of the
binding energy of a two-atomic molecule in transverse excited
state~\cite{Haller2010}, has stimulated different theoretical
modeling~\cite{Peng, Zhang, Melezhik2011, Sala2012, Sala2013}.
Attempts~\cite{Peng, Zhang} to explain this effect in one-channel
pseudopotential approach~\cite{Olshanii} were not successful.
However, a multichannel model~\cite{Melezhik2011}, including the
contribution of transversely excited channels to the confined
quasi-1D scattering process, predicts a splitting in the minimum of atomic transmission curve which is qualitatively in agreement with the observed CIR splitting~\cite{Haller2010}. Alternatively, a quantitative agreement was obtained in~\cite{Sala2012} where Sala with
co-authors linked the experimentally observed
effect with the possibility of two-atomic molecule formation in
the center-of-mass (CM) excited states when the trap is assumed to be anharmonic. In this kind of traps, pair atomic collisions can lead to resonant molecule formation in the CM
excited states~\cite{Bolda2005,Melezhik2009} due to the coupling between the relative
and CM motions . The
model~\cite{Sala2012} which was dedicated to anharmonic traps was confirmed afterwards in a special experiment for the Li molecule formation in slightly anisotropic
and anharmonic traps~\cite{Sala2013}, but the observed splitting of the
transmission due to the anisotropy of harmonic
traps~\cite{Melezhik2011} still demands theoretical and direct
experimental investigations. Therefore, one of the goals of the
present work was more profound theoretical study of the multichannel
atomic scattering confined in harmonic but anisotropic traps by considering more realistic interatomic potentials of different effective ranges.
\\
For this aim, we applied here the general grid method suggested in
\cite{Saeidian} to investigate multi-channel scattering of
identical atoms confined in a transverse harmonic trap. We suggest
a modification which makes the method applicable for anisotropic
harmonic waveguides. In \cite{Melezhik2011}, an interatomic
potential was taken in the form $V(r)=-V_{0} \exp(-{r^2}/{r_0^2})$
with rather long-range interaction due to chose of rather large
scaling parameter $r_0 \approx 5 \bar{a}$, where $\bar{a}$  is the
appropriate unit of length for the problem. Here, we study the
influence of the form of the interaction potential on
multi-channel elastic scattering inside an anisotropic harmonic
waveguide. A spacial attention was payed to investigate how the
CIR can be affected by the form of the interparticle interaction
and the strength of trap anisotropy.
\\
Ultracold molecular effects have also attracted a great interest
and in order to control them more effectively, we have to
understand the reactive collisions of ultracold atoms
\cite{Carr2009}. Chemical reactions at extremely low temperatures
were studied both experimentally \cite{Ni2010, Ospelkaus2010} and
theoretically~\cite{Idziaszek2010} using a quantum defect model
\cite{Seaton, Greene, Mies}. Idziaszek et. al
\cite{Jachymski2013,Idziaszek2015} reported analytic formulas for
reactive collision rates inside an isotropic harmonic waveguide.
Here, employing our approach we investigate the reactive rates in
the presence of an isotropic as well as anisotropic transversal
confinement in the case of one- and multi-mode regime, describing
the reactive or inelastic part of the short-range interaction as
an absorbing potential $-iV_i \exp(-{{r^2}/{r_i^2}})$ which we
have added to a $C_{12}-C_{6}$ Van der Waals interaction
potential.
\section{Multichannel Scattering Problem in an Anisotropic Harmonic Waveguide}
We study the collision of two identical bosonic atoms under the transverse anisotropic harmonic confinement defined by frequencies $\omega_{x}$ and $\omega_{y}$ along the $x$ and $y$ directions, respectively. The corresponding Hamiltonian is given by

\begin{eqnarray}
H=\sum_{i=1}^{2}(-\frac{\hbar^{2}}{2m}\nabla_{i}^2+\frac{1}{2}m\omega_{x}^2x_{i}^2+\frac{1}{2}m\omega_{y}^2y_{i}^2)+V(|{\bf r}_{1}-{\bf r}_{2}|)
\end{eqnarray}

where $m$ is the mass of the atoms and $V(|{\bf r}_{1}-{\bf r}_{2}|)$ describes the interaction potential between atoms. The Hamiltonian permits the separation of the center-of-mass ${\bf r}_{cm}=({\bf r}_1+{\bf r}_2)/2$ and relative ${\bf r}={\bf r}_1-{\bf r}_2$ variables

\begin{eqnarray}
H=H_{cm}+H_{rel}
\end{eqnarray}
where
\begin{eqnarray}
H_{cm}=-\frac{\hbar^{2}}{2M}\nabla_{cm}^2+\frac{1}{2}M\omega_{x}^2x_{cm}^2+\frac{1}{2}M\omega_{y}^2y_{cm}^2
\end{eqnarray}
and
\begin{eqnarray}
H_{rel}=-\frac{\hbar^{2}}{2\mu}\nabla_{r}^2+\frac{1}{2}\mu\omega_{x}^2x^2+\frac{1}{2}\mu\omega_{y}^2y^2+V(r)
\end{eqnarray}

Here $M=2m$ and $\mu=m/2$ are the total and reduced masses, respectively. So, the problem is reduced to scattering of a single particle with the mass $\mu$ off a central potential $V(r)$, under a transverse anisotropic harmonic confinement
\begin{eqnarray}\label{5}
\left[ -\frac{\hbar^{2}}{2\mu}\nabla_{r}^2+\frac{1}{2}\mu\omega_{x}^2x^2+\frac{1}{2}\mu\omega_{y}^2y^2+V(r) \right]\psi({\bf r})=E \psi({\bf r})
\end{eqnarray}
Here the energy of the relative two-body motion $E=E_{\perp}+E_{\parallel}$ is a sum of the transverse $E_{\perp}$ and longitudinal $E_{\parallel}={\hbar^2{k_{\parallel}}^2}/{(2\mu)}$ collision energy.\\
We assume that the interaction potential has the Gaussian
\begin{eqnarray}
V(r)=-V_{0}\exp(-\frac{r^{2}}{r_{0}^{2}})
\end{eqnarray}
or Van der Waals ($C_{12}-C_{6}$)
\begin{eqnarray}
V(r)=\frac{C_{12}}{r^{12}}-\frac{C_6}{r^6}
\end{eqnarray}
form, where $V_0$ and $r_0>0$ show the depth and the screening length of the Gaussian potential, respectively.
\\
It is more convenient to rewrite Eq.(\ref{5}) in the following rescaled form
\begin{eqnarray}\label{5a}
\left[ -\frac{1}{2}\nabla_{r}^2+\frac{1}{2}\omega_{x}^2x^2+\frac{1}{2}\omega_{y}^2y^2+V(r) \right]\psi({\bf r})=E \psi({\bf r})
\end{eqnarray}
with the scale transformation $r\rightarrow {r}/{\bar{a}}$ ($r_0\rightarrow {r_0}/{\bar a}$), $E\rightarrow {E}/{E_0}$, $\omega_{x(y)} \rightarrow {\omega_{x(y)}}/{\omega_0}$, $V\rightarrow V/{E_0}$, where $E_0={\hbar ^2}/({\mu \bar{a}^2})$ , $\omega_0={E_0}/{\hbar}$ , and $\bar{a}=4\pi {\Gamma({1}/{4})}^{-2}R_{vdW}$. Here $\Gamma(x)$ shows the gamma function and $R_{vdW}=1/2{({2\mu C_6}/\hbar^2)}^{1/4}$ stands for the Van der Waals radius~\cite{Rev2006}.
\\
In the asymptotic region $\left| z \right|\rightarrow\infty$, the transverse trapping potential dominates and the symmetrical (with respect to reflection $\bf{r}\rightarrow -\bf{r}$) wave function for the bosonic collision can be written as
\begin{eqnarray}\label{7}
\psi_{n}(\bf{r })&=&\cos(k_{n}z)\phi_{n}(x,y)\nn\\
&&+\sum_{{n}^{\prime}}f_{nn^{\prime}}\exp(ik_{{n}^{\prime}}\left| z \right|)\phi_{{n}^{\prime}}(x,y)
\end{eqnarray}

where

\begin{eqnarray}
k_{n}=\sqrt{2(E-E_{\perp}^{n})}
\end{eqnarray}
\\
and
\begin{eqnarray}
\phi_{n}(x,y)=\phi_{n_{x},n_{y}}(x,y)=\phi_{n_{x}}(x)\phi_{n_{y}}(y)
\end{eqnarray}
is the eigenfunction of the 2D harmonic oscillator, corresponding to the transversal eigenenergy
\begin{eqnarray}
E_{\perp}^{n}=E_{\perp}^{n_{x},n_{y}}=\left[ \omega_{x}(n_{x}+\frac{1}{2})+\omega_{y}(n_{y}+\frac{1}{2}) \right]
\end{eqnarray}
$\phi_{n_{x}}(x) (\phi_{n_{y}}(y))$ being the eigenfunction of a 1D harmonic oscillator along the $x (y)$ axis.
The scattering amplitude $f_{n{n}^{\prime}}$ describes transition from the initial $n=(n_{x},n_{y})$ to the final $n^{\prime}=({n_{x}}^{\prime},{n_{y}}^{\prime})$ transverse state. The summation is over all open channels.
\\
The transmission probability from the initial transverse state $n$ to all possible final open states ${n}^{\prime}$ in the course of the collision of identical bosons is defined as
\begin{eqnarray} \label{13}
T_{n_{x}n_{y}}=T_n=\sum_{{n}^{\prime}}\frac{k_{{n}^{\prime}}}{k_{n}}{\left| \delta_{n{n}^{\prime}}+f_{n{n}^{\prime}} \right|}^2\,\,.
\end{eqnarray}
The scattering amplitude is related to the 1D S matrix via $S_{nn'}=\delta _{nn'}+2f_{nn'}$. At zero-energy limit, it is convenient to parameterize the  1D scattering through the 1D scattering length as $a_{1D}=i(1+f_{00})/(k_0f_{00})$ \cite{Olshanii}, where $f_{00}$ shows the scattering amplitude from the initial $n=0=(0,0)$ state to the same final state $n^{\prime}=0=(0,0)$.
\\
If there is an absorbing potential in short range, then the corresponding $a_{1D}$ becomes a complex value with the imaginary part related to inelastic processes (reactions). The loss mechanism characterised quantitatively by the reactive rate constant $K^{1D,re}$. In one dimension, the reactive and elastic collision rates for identical particles in channel $n$ are defined as~\cite{Simoni2015}
\begin{eqnarray}
K^{1D,re}&=&{k_n}\left( 1-{\left| S_{nn} \right|}^2 \right) \label{14}\\
K^{1D,el}&=&{k_n}{\left|1-S_{nn} \right|}^2
\end{eqnarray}
\section{Numerical Approach}
To obtain the transmission probability and the rate constants, we
have to calculate the scattering amplitude
$f_{n{n}^{\prime}}$ by
integration of the Schr\"odinger Eq.(\ref{5a}) with the boundary
asymptotic condition Eq.(\ref{7}). Due to the anisotropy of the
confinement, the scattering problem is non-separable in the three
dimensional space $(r,\theta,\phi)$. To resolve this problem we
use the discrete variable method suggested in \cite{Melezhik91,Melezhik2003}.
\\
First, we discretize the Schr\"odinger Eq.(\ref{5a}) on a 2D grid
over $(\theta,\phi)$ variables according to the non-direct product
DVR~\cite{Melezhik1997,Melezhik99,Melezhik2016}. The idea of this approach is in
expanding the desired wave function $\psi(r,\theta,\phi)$ in the
basis orthogonal and complete on the 2D grid
$\Omega_j=(\theta_j,\phi_j)$

\begin{eqnarray}
f_{j}(\Omega)\approx\sum_{\nu=1}^{N}\bar{Y}_{\nu}(\Omega)\left[
\mathbf{Y}^{-1}\right]_{\nu j}
\end{eqnarray}
so that

\begin{eqnarray}\label{15}
\psi(r,\theta,\phi)\approx\frac{1}{r}\sum_{j=1}^{N}f_{j}(\Omega)u_{j}(r)\,\,.
\end{eqnarray}

Here $N=N_{\theta}\times N_{\phi}$, where $N_{\theta}$ and
$N_{\phi}$ are the number of grid points over the $\theta$ and
$\phi$ variables respectively. The $N\times N$ matrix
$\mathbf{Y}^{-1}$ is inverse to  the matrix $\mathbf{Y}$ defined
as $Y_{j\nu}=\sqrt{\lambda_{j}} \bar{Y}_{\nu}(\Omega_j)$, with
$\lambda_j=({2\pi \lambda_j^{\prime}})/{N_{\phi}}$ and
$\lambda_j^{\prime}$ being the weights of the Gaussian quadrature
over $\theta$. The angular grid points $\theta_{j}$ and $\phi_j$
are defined as the zeros of the Legendre polynomial
$P_{N_{\theta}}(\cos \theta)$ and
$\phi_j=2\pi j/N_{\phi}$ , respectively. The symbol $\nu$ represents the two fold
index $\nu=(l,m)$ and the summation over $\nu$ is equivalent to

\begin{eqnarray}
\sum_{\nu=1}^{N}=\sum_{m=-\frac{N_{\phi}-1}{2}}^{\frac{N_{\phi}-1}{2}}\sum_{l=\left| m \right|}^{\left| m \right|+N_{\theta}-1}
\end{eqnarray}

 The polynomials $\bar{Y}_{\nu}(\Omega)$ are chosen as
 \begin{eqnarray}
 \bar{Y}_{\nu}(\Omega)=\bar{Y}_{lm}(\Omega)=e^{im\phi}\sum_{l^\prime}c_{l}^{l^\prime}P_{l^\prime}^m(\theta)
 \end{eqnarray}

 where $c_{l}^{l^\prime}=\delta_{ll^\prime}$ holds in general, thus $\bar{Y}_{\nu}(\Omega)$ coincide with the usual spherical harmonics $Y_{\nu}(\Omega)$
 except some higher values of $\nu$ so that the orthogonality relation remains

 \begin{eqnarray}
 \langle \bar{ Y}_{\nu} \, |\bar{ Y }_{\nu ^\prime} \rangle = \int {\bar{Y_{\nu}}}^{*}(\Omega)\bar{Y_{\nu^\prime}}(\Omega)d\Omega\approx\sum_{j=1}^{N}\lambda_{j}Y_{\nu}^{*}(\Omega_j)Y_{\nu^\prime}(\Omega_j)=\delta_{\nu\nu^\prime}
\end{eqnarray}

In a few cases involving the highest $l$ values ($l\geq N_{\theta}$), $Y_{\nu}(\Omega)$ have to be orthogonalized. We address the set of orthogonal basis as $\bar{ Y }_{lm}(\Omega)$. First, for $l=N_{\theta}$ we make a polynomial orthogonal to the ones of lower $l$ value

\begin{eqnarray}
\widetilde{ Y }_{lm}(\Omega)=Y_{lm}(\Omega)-\sum_{l^\prime=\left| m \right|}^{l-1}\langle Y_{lm} \, |\bar{ Y }_{l^\prime m} \,  \rangle \bar{ Y }_{l^\prime m}(\Omega)
\end{eqnarray}

and, we make it normalized

\begin{eqnarray}
\bar{ Y }_{lm}(\Omega)=\frac{\widetilde{ Y }_{lm}(\Omega)}{\langle \widetilde{ Y }_{lm} \, |\widetilde{ Y }_{lm}\,  \rangle}
\end{eqnarray}
then, we perform the above procedure iteratively in order to
obtain $\bar{ Y }_{lm}(\Omega)$ for the next values of
$l$. Such a way, the above orthogonalization Gramm-Schmidt
procedure leads as to the basis (16) which is orthonormal  and
complete on the grid $\Omega_j$ for any chosen $N$.
\\
By substituting Eq.(\ref{15}) into Eq.(\ref{5a}), we reach to a system of $N(=N_{\theta} \times N_{\phi})$ Schr\"odinger-like coupled equations with respect to the $N-$dimensional unknown vector $\mathbf{u}(r)=\left\{ \sqrt{\lambda_{j}}u_{j}(r) \right\}_{1}^N$

\begin{eqnarray}\label{21}
\left\{ \textbf{H}^{(0)}(r)+2\left[E\textbf{I}-\textbf{V}(r) \right] \right\}\textbf{u}(r)=0
\end{eqnarray}

where \textbf{I} is the unit matrix and

\begin{eqnarray}
\textbf{H}_{jj^\prime}^{(0)}(r)=\delta_{jj^\prime}\frac{d^2 }{d r^2}-\frac{1}{r^2}\sum_{\nu=1}^{N}Y_{j\nu}l(l+1)(\mathbf{Y}^{-1})_{\nu j^\prime},\\
V_{jj^\prime}(r)=V(r,\Omega_{j})\delta_{jj^\prime}=\left\{
V(r)+\frac{1}{2}\omega_{x}^2x_{j}^2+\frac{1}{2}\omega_{y}^2y_{j}^2
\right\}\delta_{jj^\prime}
\end{eqnarray}
Here $x_{j}=r\sin \theta_{j}\cos \phi_{j}$ and $y_{j}=r\sin
\theta_{j}\sin \phi_{j}$ and the elements $u_j(r)$ of the vector
$\textbf{u}(r)$ coincide with the values $r\psi(r,\Omega_j)$ of
desired wave function on the grid points $\Omega_j$.
\\
We solve the system of Eqs. (\ref{21}) on a quasi-uniform radial grid  \cite{Melezhik1997}
\begin{eqnarray}
r_{j}=R\frac{e^{\gamma x_{j}}-1}{e^\gamma-1},~~~~~j=1,2,...,N_{r}
\end{eqnarray}
of $N_r$ grid points $\left\{r_j \right\}$ defined by mapping $r_j \in [0,R \rightarrow +\infty ]$ onto the uniform grid $x_j \in [0,1]$ with the equi-distant distribution $x_j-x_{j-1}=1/N_r$. One can achieve a suitable distribution of the grid points for a specific interatomic and confining potential by varying $N_r$ and the parameter $\gamma >0$.
\\
Thus, after the finite difference approximation, the initial 3D Schr\"odinger Eq.(\ref{5a}) is reduced to the system of $N_{r}$ algebraic matrix equations
\begin{eqnarray}\label{25}
\sum_{p=1}^{3}A_{j-p}^ju_{j-p}+\left( A_{j}^j+2\left\{ EI-V_{j}\right\}\right)u_{j}+\sum_{p=1}^{3}A_{j+p}^ju_{j+p}=0,\nn\\
j=1,2,...,N_{r}-3,\nn\\
u_{j}+\alpha_{j}^{\left(1\right)}u_{j-1}+\alpha_{j}^{\left(2\right)}u_{j-2}+\alpha_{j}^{\left(3\right)}u_{j-3}=\gamma_{j},\nn\\
j=N_{r}-2,N_{r}-1,N_{r}
\end{eqnarray}
In the first three equations, the functions $u_{-3}$, $u_{-2}$, $u_{-1}$ and $u_0$ are set to zero by using the left-side boundary condition: $u_0=0$ and $u_{-j}=-u_j$. The last equation for large values of $j$ comes from the right-side boundary
condition Eq.(\ref{7}) at large $r$ by eliminating the unknown
scattering amplitudes from Eq.(\ref{7}) written for a few largest
values of $r_j$. We solve this boundary value problem
Eq.(\ref{25}) by using the LU decomposition approach \cite{PreTeu}
or the sweep method \cite{Gelfand} which is adopted to the
multichannel atomic scattering in \cite{Saeidian} (see Appendix
A).

By mapping the calculated wave function at the points $z_j=r_j\cos
\theta_j\rightarrow \pm\infty$ with the asymptotic boundary
condition (\ref{7}) we find the scattering amplitude
$f_{n{n}^{\prime}}$.

\section{Results and Discussion}
\subsection{Multichannel elastic scattering}
\begin{figure}[H]
\begin{center}
\begin{subfigure}{
\includegraphics[width=0.5\textwidth]{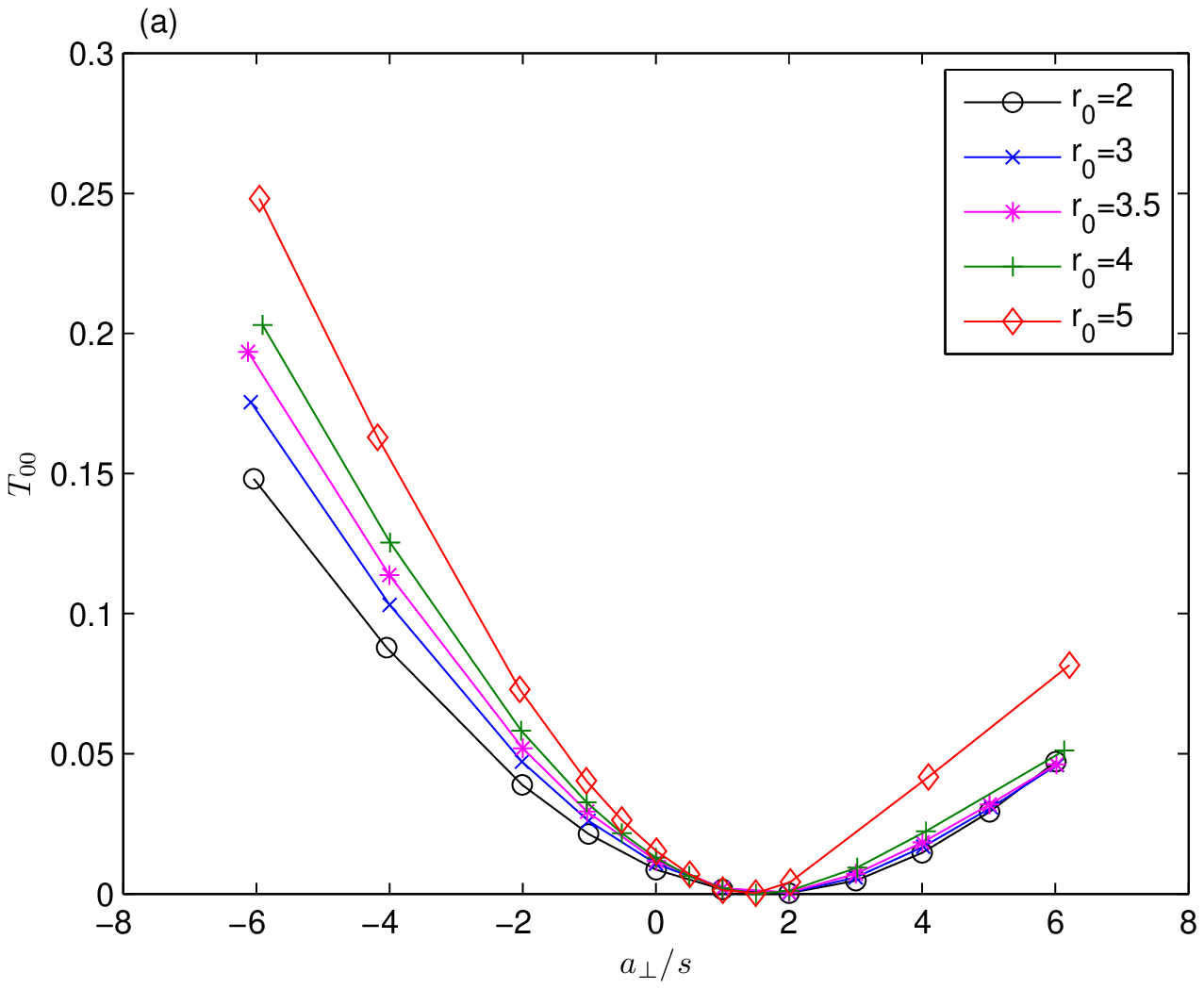}}
\end{subfigure}\\
\begin{subfigure}{
\includegraphics[width=0.5\textwidth]{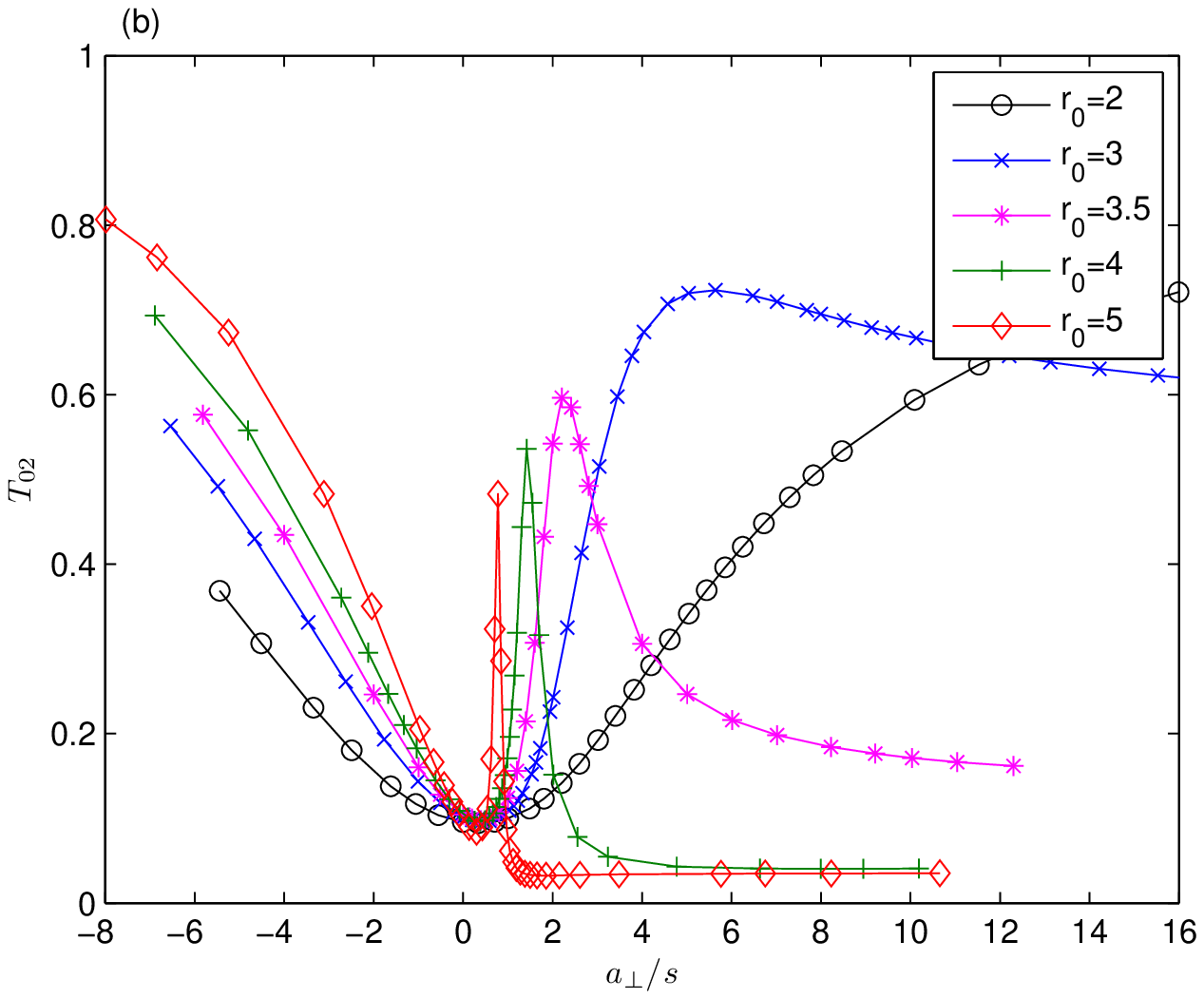}}
\end{subfigure}
\begin{subfigure}{
\includegraphics[width=0.5\textwidth]{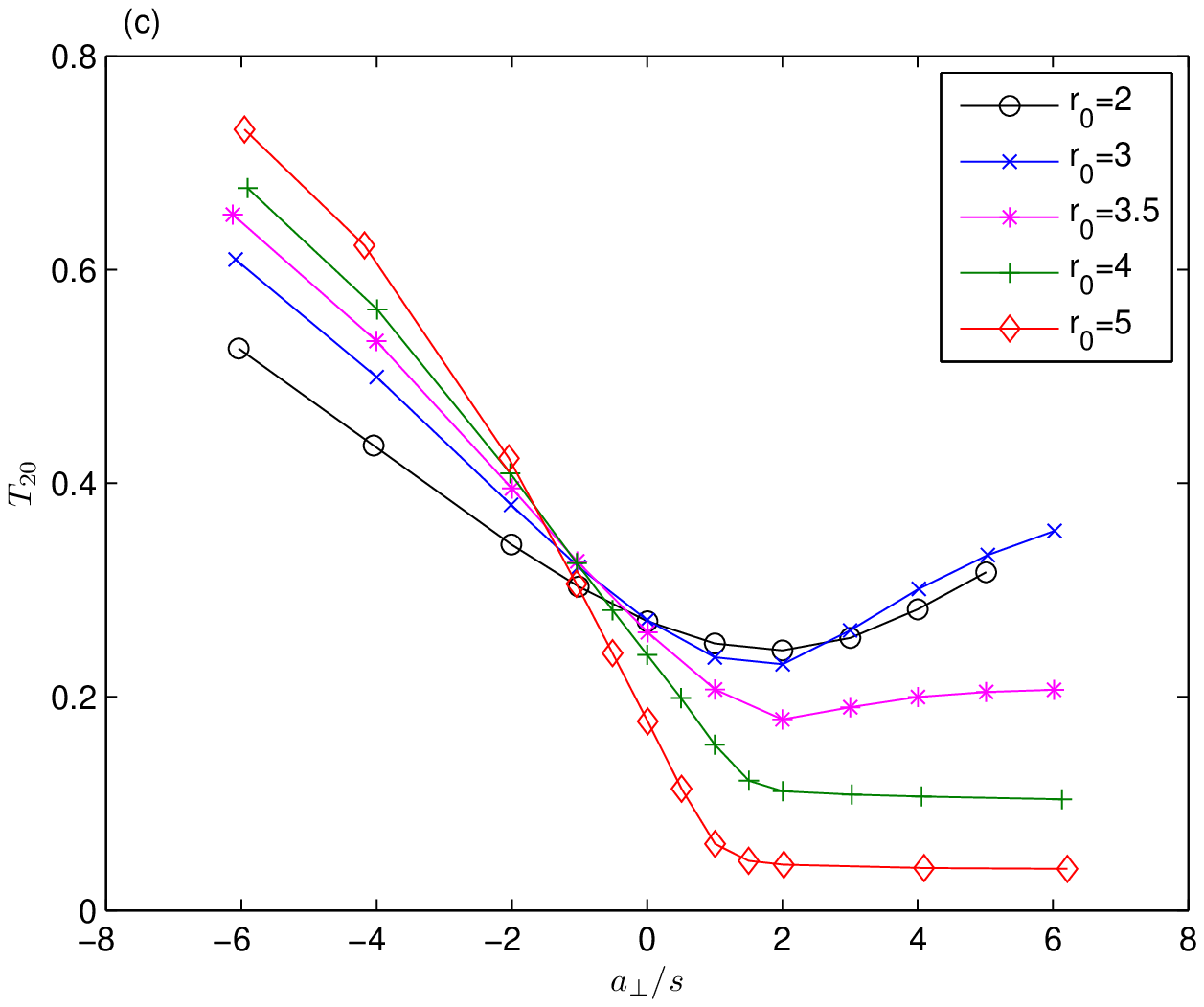}}
\end{subfigure}
\end{center}
\caption{(Color online) Partial transmission coefficients ($T_{00}$, $T_{02}$ ,$T_{20}$) for several $r_0$ as a function of ${a_{\perp}}/s$ when the anisotropy is fixed at ${\omega _{x}}/{\omega_{y}}=1.05$ ($\omega_y=0.02\omega_0$) and the initial state of the system is in the last open channel with a longitudinal energy ${E_\parallel}/{E_\perp ^0}=5\times 10^{-3}$ for the cases of (a) one open channel (b)two open channels (c) three open channels.}
\label{fig:Fig1}
\end{figure}

In this Subsection we analyze pair atomic collisions in anisotropic waveguides at colliding energies permitting  transition between the first three transverse channels with $n=(n_x n_y)=(00),(02)$ and $(20)$. We investigate the influence of the form of interatomic interaction $V(r)$ and the strength of the anisotropy $1-\omega_x/\omega_y$ on the transmission coefficients $T_n(E,\omega_x/\omega_y)$ (13). Special attention is paid on what found in Ref.\cite{Melezhik2011}, i.e. splitting of the transmission coefficient $T_{02}(E,\omega_x/\omega_y)$ in the case $\omega_x/\omega_y\neq 1$ which can be responsible for the CIR splitting experimentally observed in the ultracold gas of Cs atoms confined in an anisotropic waveguide \cite{Haller2010}.

First, the interaction potential $V(r)$ was chosen in the Gaussian form $V(r)=-V_{0} \exp(-r^2/r_0^2)$ suggested in~\cite{Melezhik2011} for the fixed anisotropic waveguide with $\omega _{x}/\omega_{y}=1.05$ ($\omega_y=0.02 \omega_0$) and longitudinal colliding energies $E_\parallel/E_\perp ^0=5\times 10^{-3}$ which reckoned from the threshold of the last open channel. Calculations were performed with the varying parameters $V_0$ and $r_0$ which supplyed entering to the area near the point of CIR. As known, CIR was predicted~\cite{Olshanii} and experimentally observed~\cite{Haller2010} for an isotropic waveguide ($\omega_{\perp}=\omega_x=\omega_y$) at the point $a_{\perp}/s=1.46...$, i.e. when the dimensionless s-wave scattering length $s=-\lim_{k\rightarrow 0} {\delta_0(k)}/k$ approaches to the value of the trap width $a_{\perp}={1}/{\sqrt{\omega_{\perp}}}$. Here $\delta_0(k)$ represents the energy dependent phase shift due to the scattering in free space. Therefore, by integrating the scattering problem (8,9) for various depth $V_0$ and widths $r_0$ of the potential $V(r)$ and fixed $\omega_{\perp}$ and $E$, we get the transmission coefficients $T_{n_x n_y}(E,a_{\perp}/s,r_0)$ and find the CIR position where $\Re f_{00}(E,\omega_{\perp},V_0,r_0) \rightarrow -1$, $\Im f_{00}(E,\omega_{\perp},V_0,r_0)\rightarrow 0$ and $T_{00}(E,\omega_{\perp},V_0,r_0)\rightarrow 0$ (see definition CIR in ~\cite{Olshanii}). With the fixed parameters $V_0$ and $r_0$ the problem (8) is integrated for $\omega_{\perp}=0$ as well, and the scattering amplitude $f_0$ is extracted from the free-space scattering asymptotic. Such a way we calculate the scattering length corresponding to the chosen $V_0$ and $r_0$. In \figref{Fig1}(a) ( (b), (c)), we present the partial transmission coefficients for the cases when one (two, three) channel(s) are open, and the colliding atoms are initially in the last open channel. Rather smooth dependence of the transmission coefficients $T_{00}$ and $T_{20}$ on the ratio $a_{\perp}/s$ and $r_0$ near CIR is clear from \figref{Fig1}(a) and (b). However, the minimum of the coefficient $T_{02}$ at CIR splits with increasing $r_0$. The effect becomes observable for anisotropy $\omega_y/\omega_x=1.05$ at $r_0\geq 3.5$. So, we can conclude that the CIR splitting occurs only for rather long-range interaction considerably overlapping with the harmonic confining potential. The effect was already observed in~\cite{Melezhik2011} where the calculations of the transmission coefficients where performed with the same Gaussian potential (6) but for only one fixed parameter $r_0=5 \bar{a}$.

\begin{figure}[H]
\centering\includegraphics[width=\textwidth]{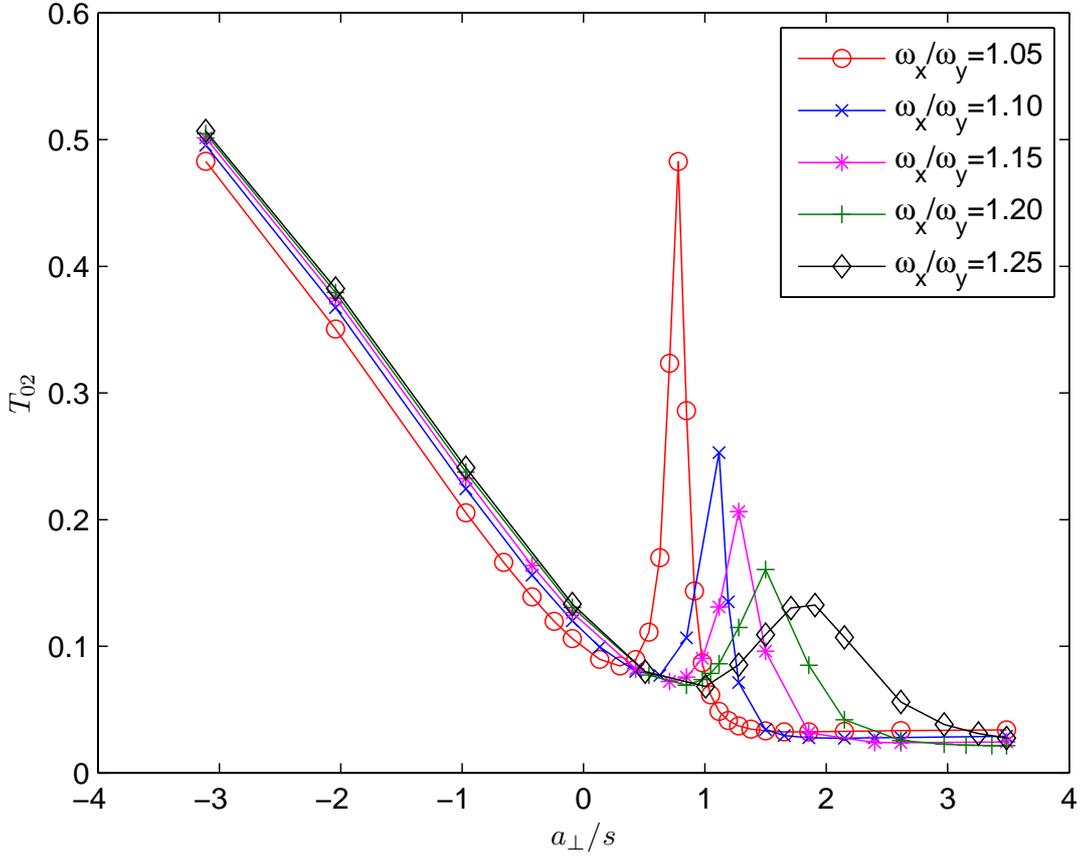}
\caption{ (Color online) Partial transmission coefficient $T_{02}$ for various anisotropies as a function of $a_{\perp}/s$ when $r_0=5 \bar{a}$ , ${E_\parallel}/{E_\perp ^0}=5\times 10^{-3}$ and $\omega _y=0.02 \omega _0$. }
\label{fig:Fig2}
\end{figure}

\figref{Fig2} illustrates the dependence of the value of the CIR splitting on the  strength of the anisotropy by assuming a long-range character ($r_0=5 \bar{a}$) of the interaction potential $V(r)$ (6). As the anisotropy increases, the distance between the two minima in $T_{02}(a_{\perp}/s)$ increases, as well. The same phenomenon is observed in the experimental result \cite{Haller2010}, which is also in quantitative agreement with the theoretical modeling in \cite{Sala2012}. Squeezing or stretching the confinement along one axis while keeping it constant along the other axis, leads to elimination of the degeneracy between the eigenstates $(n_x,n_y)=(0,2)$ and $(2,0)$ of the transverse confining potential and increasing the energy splitting between the energy thresholds of these two scattering channels, and hence larger CIR splitting.
\begin{figure}[H]
\centering\includegraphics[width=\textwidth]{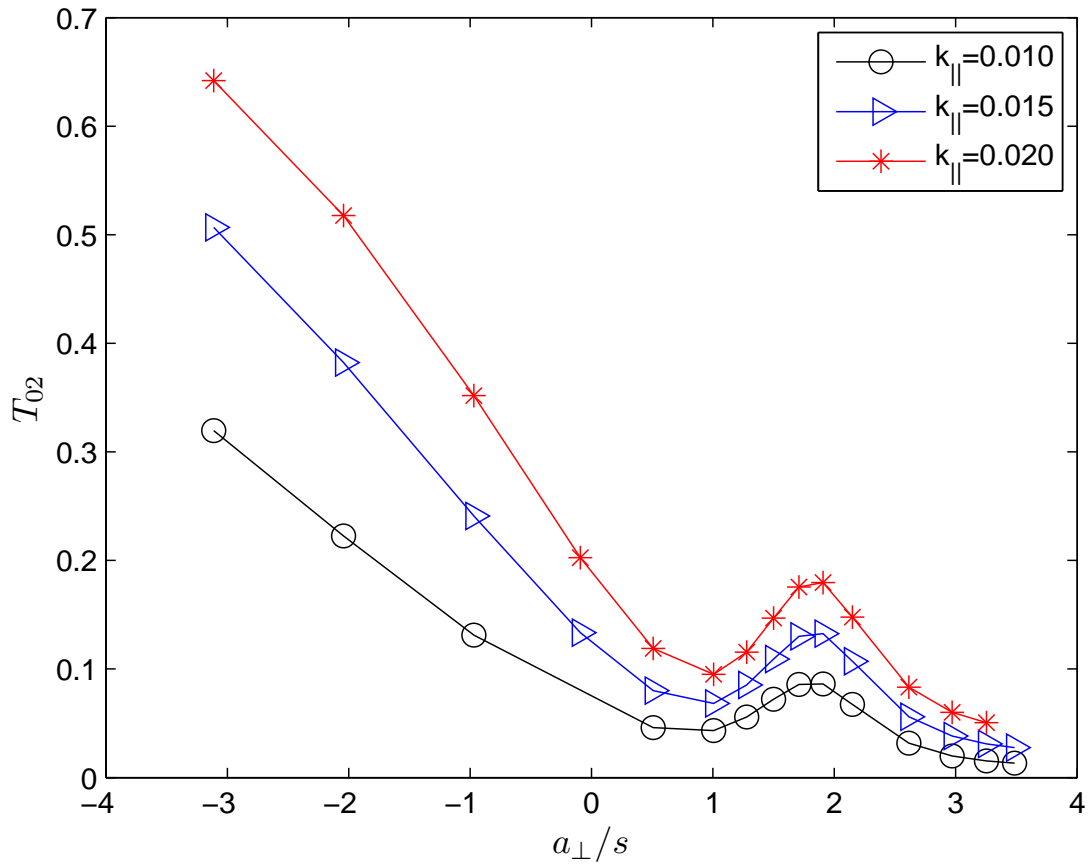}
\caption{ (color online) The dependence of the partial transmission coefficient $T_{02}$ on the longitudinal energy for ${\omega _{x}/}{\omega_{y}}=1.25$ ($\omega _y=0.02 \omega _0$) and $r_0=5 \bar{a}$ as a function of $a_{\perp}/s$.}
\label{fig:Fig3}
\end{figure}
In \figref{Fig3} we can see the effect of longitudinal energy $E_{\parallel}$ on the behaviour of transmission curve $T_{02}(E,a_{\perp}/s)$  for a considerable anisotropy in transversal confinement ($\frac{\omega _{x}}{\omega_{y}}=1.25$). It shows that the splitting becomes more pronounced with increasing energy.
\begin{figure}[H]
\centering\includegraphics[width=\textwidth]{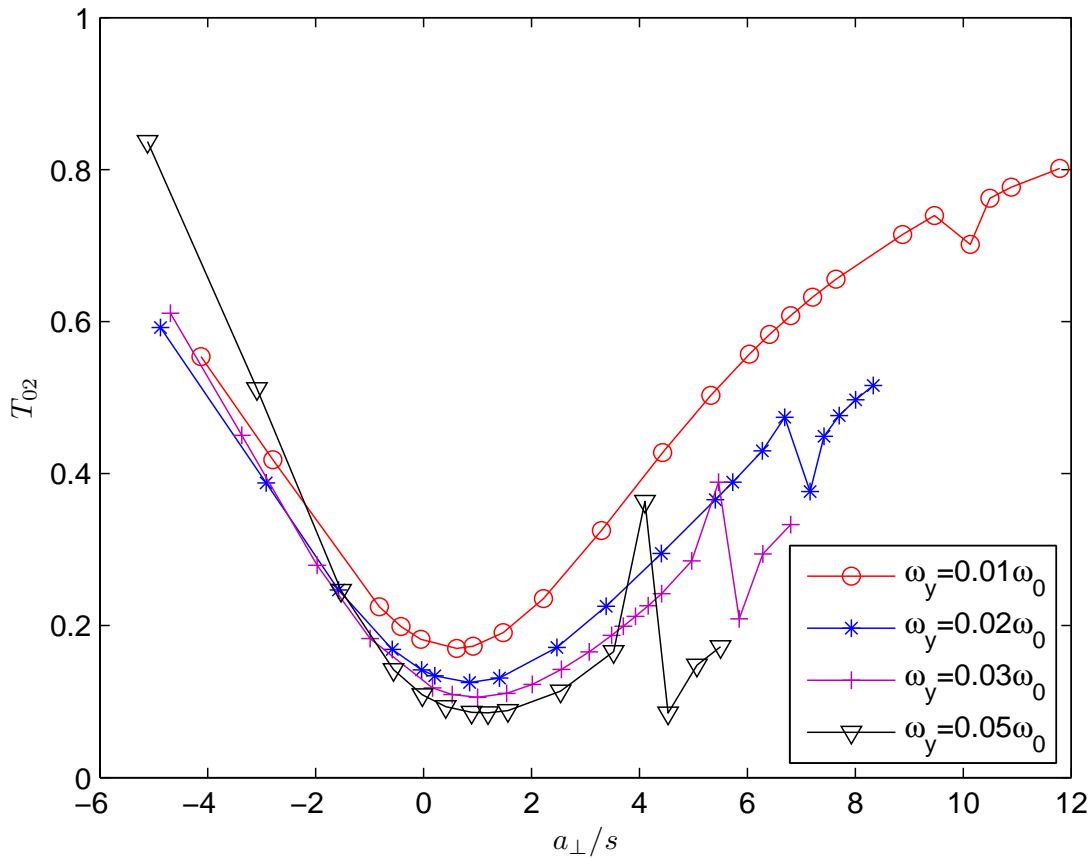}
\caption{(color online) Transmission coefficient $T_{02}$, considering the van der Waals interacting potential, as a function of $a_{\perp}/s$ for various trap frequencies along the $y$ axis, when the anisotropy is fixed at ${\omega _x}/{\omega _y}=1.1$ and the longitudinal wave vector equals to $k_\parallel=0.02$ .}
\label{fig:Fig4}
\end{figure}
We have also analyzed the problem with more realistic $C_{12}-C_{6}$ interaction, i.e.
van der Waals potential. For the pair
of $^{133}{Cs}$ atoms the parameters of the van der Waals potential are $C_6=6890~a.u.$ ,  $R_{vdW}=101a_0$ ,
$\bar{a}=96.55 a_0$ (where $a_0$ is the Bohr radius). In this case, the units of the problem becaomes $E_0=8.8\times
10^{-10}E_h$ ,and $\omega_0=2\pi \times 5.7 \times 10^3 kHz$. In \figref{Fig4}
we present the calculated curve $T_{02}(E,a_{\perp}/s)$ of the transmission coefficient for a fixed value of
${\omega _{x}}/{\omega_{y}}=1.1$. As it can be seen, only for very
tight traps, the minimum is splitted. In this respect, one can evaluate the van der Waals potential
as a very short-range interaction that can lead to the CIR splitting only in the case of its
overlap with a very tight harmonic confinement. The form of the splitting in the case of $C_{12}-C_6$ interaction (7) is considerably different from the case of Gaussian interaction (6).

\subsection{Inelastic scattering}
In this Subsection we analyze the case where chemical reactions
occur at short distances that leads to inelastic scattering.
Here we consider collision of two particles interacting via the
$C_{12}-C_6$ van der Waals potential plus an imaginary term like $-i V_i
\exp(-{r^2}/{r_i^2})$ at short distances of the order $r_i$, which
models the loss mechanism due to presence of the inelastic channel
of a molecule formation. This process happens at distances much smaller than the length scale associated with the long range Van der Waals potential, $\bar{a}$. Here we have chosen
$r_i=0.4 \bar{a}$ in order
to reproduce the complex scattering length presented in
\cite{Idziaszek2010} where they used this value to compute the s-wave reactive collision rate for $^{40}{K} ^{87}{Rb}$  molecules and found it compatible with the corresponding
experimental data on \cite{Ospelkaus2010}.
\\
The scattering process can be parametrized by using two dimensionless quantum defect parameters $0\leqslant y \leqslant 1$ and $s$ , where $y$ (determined here by $V_i$ and $r_i$) is related to the short range reaction probability via $P^{re}={4y}/{(1+y)^2}$ \cite{Jachymski2013} and $s$ represents the dimensionless 3D scattering length in the absence of any loss process. In general, the phase shift $\delta_0$ of the s-wave scattering process in free space becomes a complex number $\tilde{\delta}_0$ leading to the complex scattering length $\tilde{ a }=\alpha-i \beta$ defined as $\tilde{ a }=-\lim_{k\rightarrow 0} {\tilde{\delta}_0(k)}/k$.
\begin{figure}[H]
\centering\includegraphics[width=\textwidth]{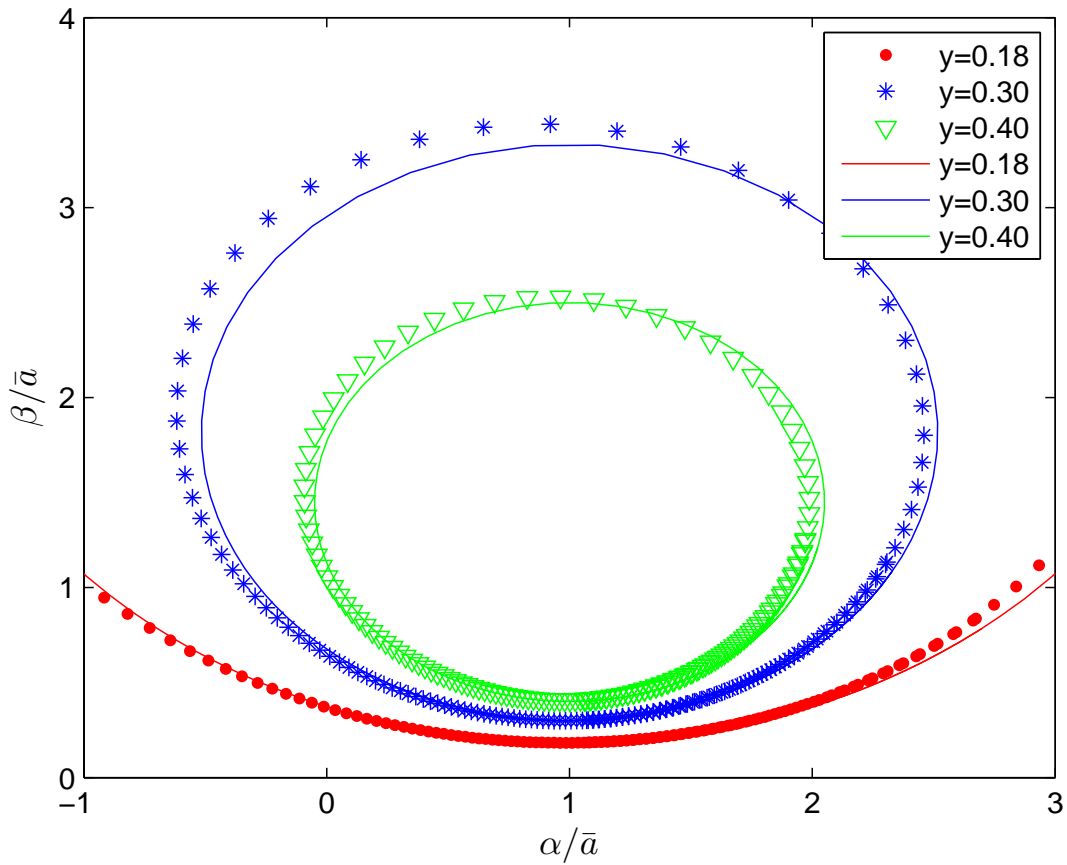}
\caption{ (Color online) Real and imaginary parts of the scattering length in free
space with absorption as $k\rightarrow0$ and $s$ varies over its
full range $-\infty<s<\infty$ . Analytical (numerical) values for
different values of the loss parameter $y$ are shown with
curves (dots).} \label{fig:Fig5}
\end{figure}
In \figref{Fig5} we have plotted complex scattering length
($\tilde{ a }=\alpha-i \beta$) in free space obtained at the
numerical integration of Eq.(\ref{5a}) with $\omega_x=\omega_y=0$ along
with the corresponding analytical values for comparison. Our
results are in a good agreement with the analytical results which at the low energy limit is a function of $y$ and $s$ (in units of $\bar{a}$) \cite{Idziaszek2010}
\begin{eqnarray}
\tilde{ a }=s+y \frac{1+{(1-s)}^2}{i+y(1-s)}
\end{eqnarray}
In \figref{Fig6} we present the calculated real and imaginary
parts of the complex scattering length $\tilde{ a }=\alpha-i
\beta=-{1}/{a_{1D}}$ inside an isotropic harmonic waveguide for
various loss parameters, where $a_{1D}=i{k_0f_{00}}/{(1+f_{00})}$ is the quasi-1D scattering length (see the definition after Eq.(\ref{13})).  When there is no absorption
($y=0$), $\alpha$ exhibits a resonance at $a_{\perp}/s=1.46$. But in
the presence of absorption ($y\neq 0$), the behaviour of $\alpha$ changes dramatically and there is no divergence at $a_{\perp}/s=1.46$, hence the resonance becomes less pronounced and disappears gradually by incresing $y$. The
strongest loss occurs at the maximum value of $\beta$ which is
located at the CIR position.
\begin{figure}[H]
\begin{center}
\begin{subfigure}{
\includegraphics[width=0.8\columnwidth]{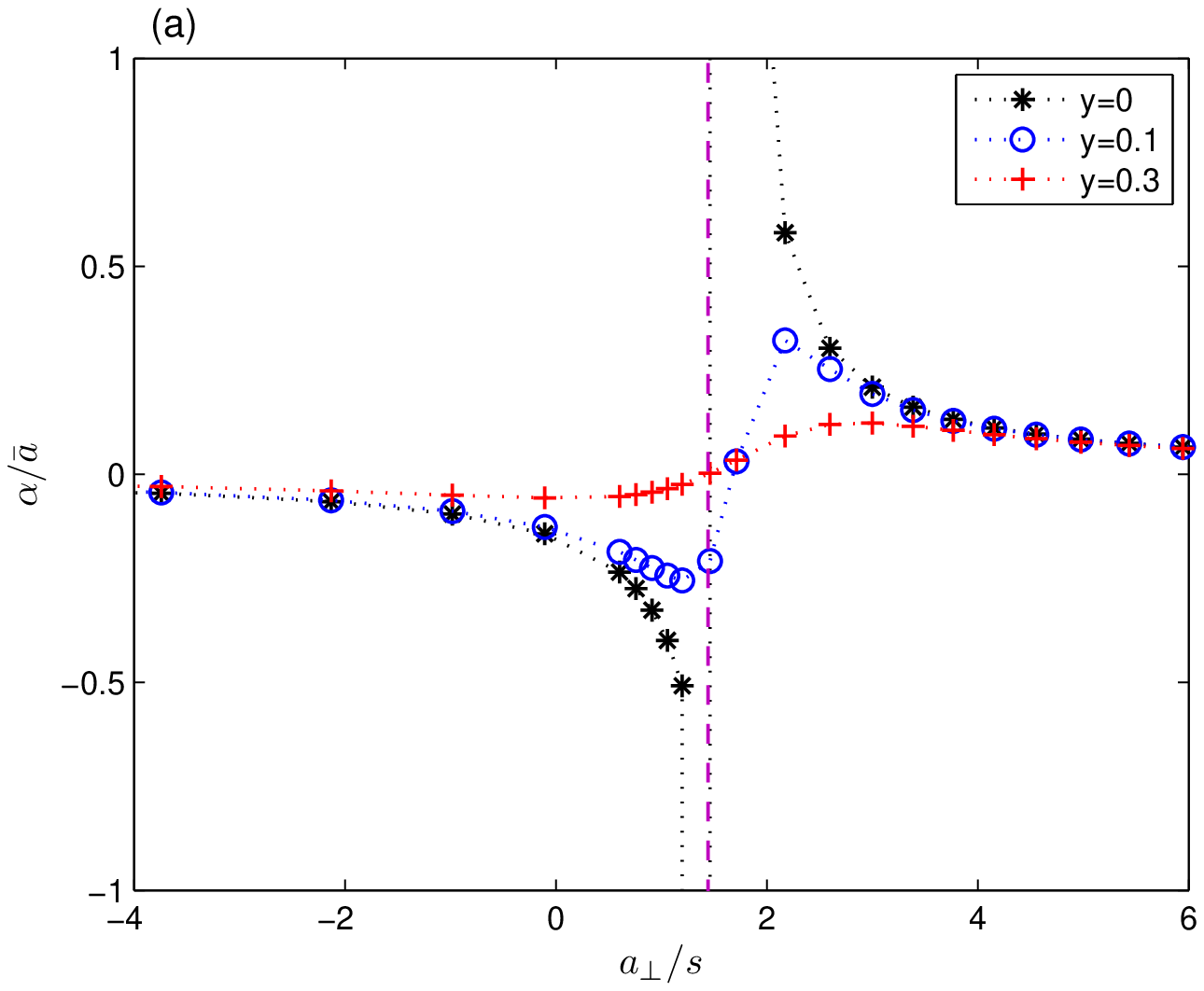}}
\end{subfigure}\\
\begin{subfigure}{
\includegraphics[width=0.8\columnwidth]{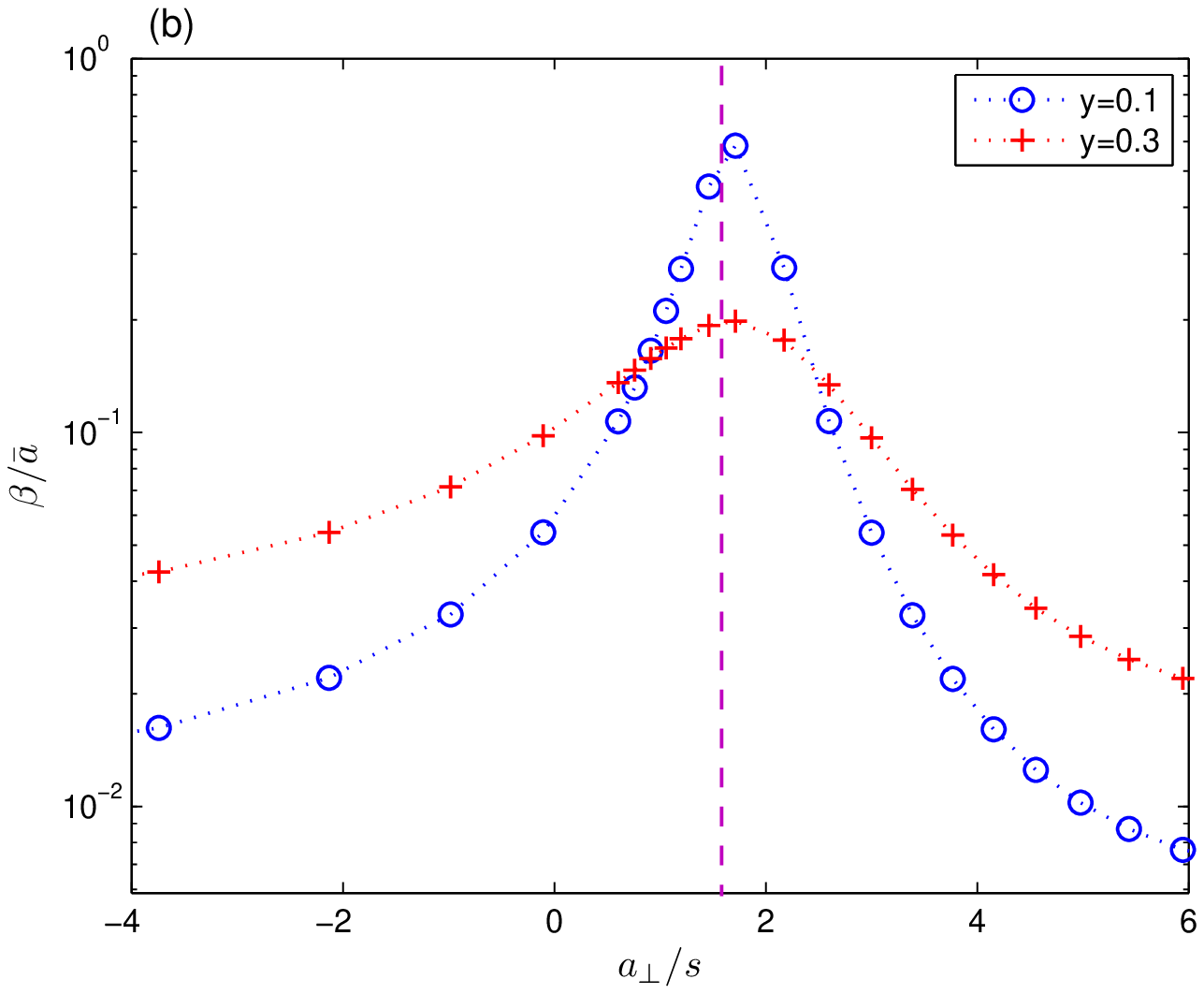}}
\end{subfigure}
\end{center}
\caption{ (Color online)(a)real and (b)imaginary part of the complex scattering length inside an isotropic harmonic waveguide with frequency $\omega_{\perp}=0.02\omega_0$ as a function of $a_{\perp}/s$ when $k_0 \rightarrow0$ for different loss parameters. The dashed line shows the position of CIR in the absence of absorption ($a_{\perp}/s=1.46$).}
\label{fig:Fig6}
\end{figure}

\begin{figure}[H]
\centering\includegraphics[width=\textwidth]{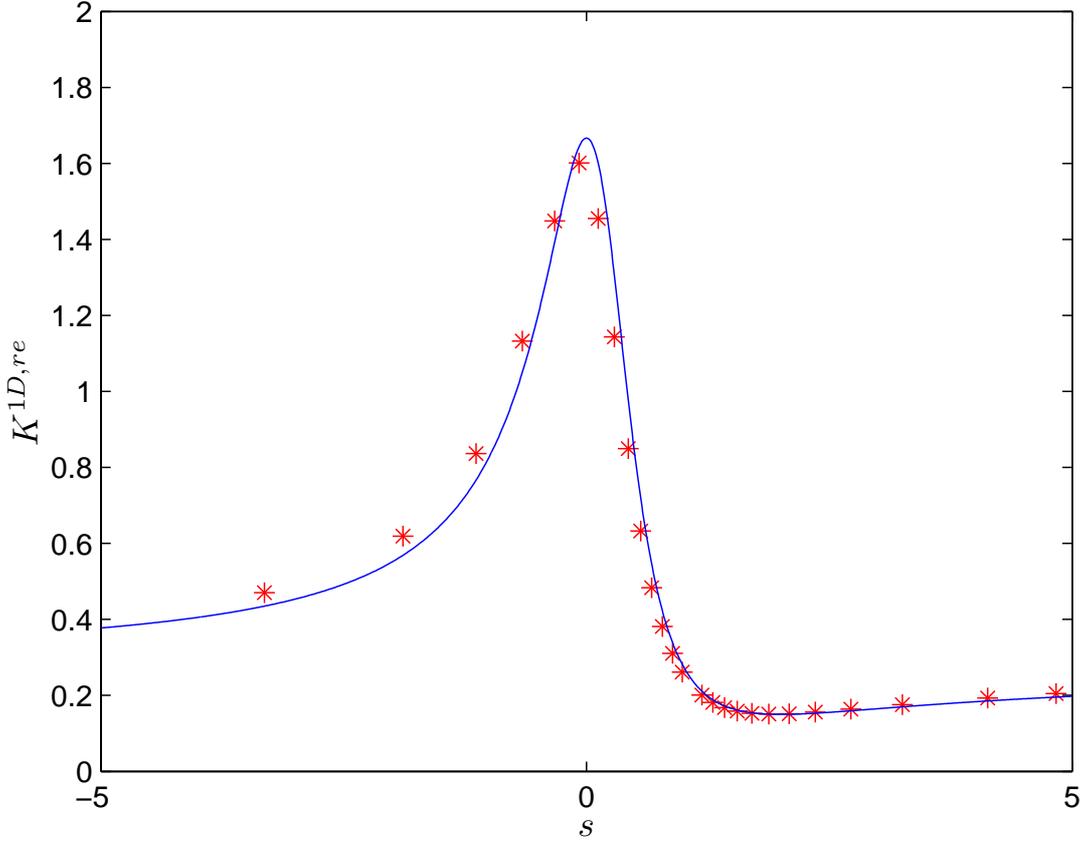}
\caption{(Color online)The blue curve (red dots) show(s) the dimensionless reactive rate constant at very low collision energies in an isotropic harmonic trap with frequency $\omega_{\perp} =0.02 \omega_0$ obtained from our calculations (Eq.(\ref{re_K}) ) for a loss parameter $y=0.3$.}
\label{fig:Fig7}
\end{figure}

\begin{figure}[H]
\begin{center}
\begin{subfigure}{
\includegraphics[width=0.7\columnwidth]{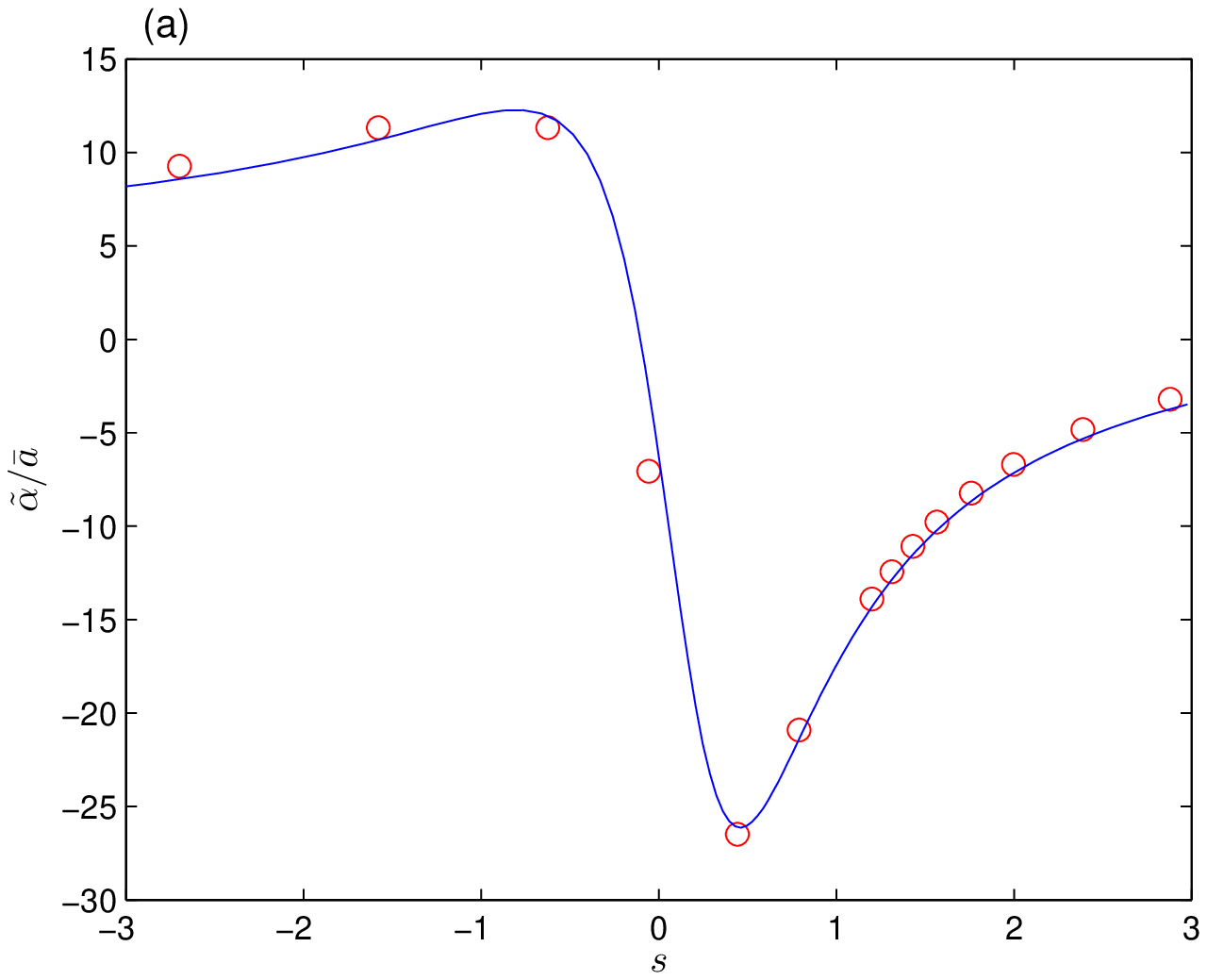}}
\end{subfigure}\\
\begin{subfigure}{
\includegraphics[width=0.7\columnwidth]{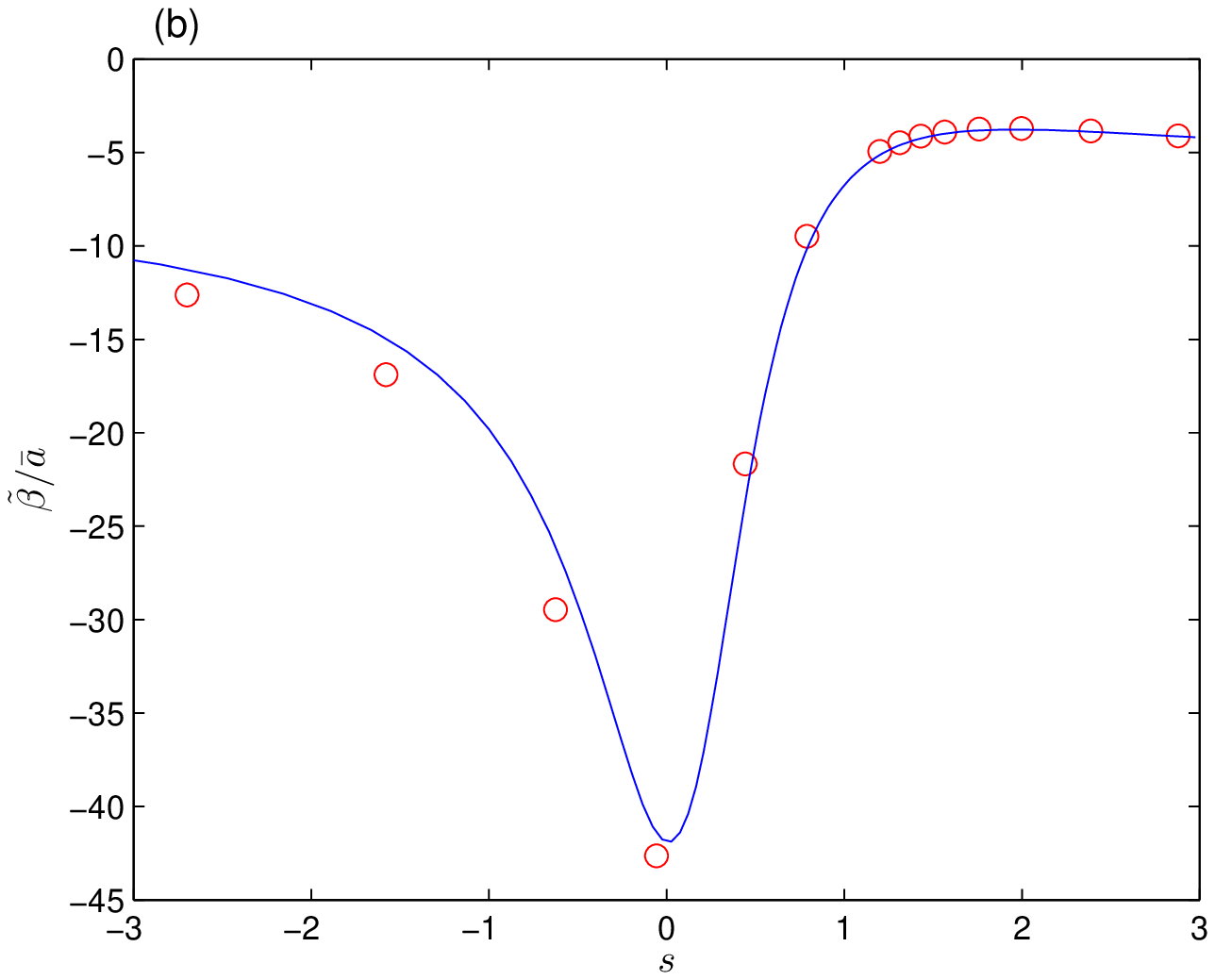}}
\end{subfigure}
\end{center}
\caption{ (Color online)(a)Real and (b)imaginary part of the 1D scattering length
$a_{1D}$ as a function of 3D scattering length $s$ when
$k_0 \rightarrow0$ for $y=0.3$, $\omega_x/\omega_y=1.05$
and $\omega_y=0.02 \omega_0$. Dots (curves) correspond to the
numerical (analytical \cite{Peng}) results.}
\label{fig:Fig8}
\end{figure}

\begin{figure}[H]
\begin{center}
\begin{subfigure}{
\includegraphics[width=0.7\columnwidth]{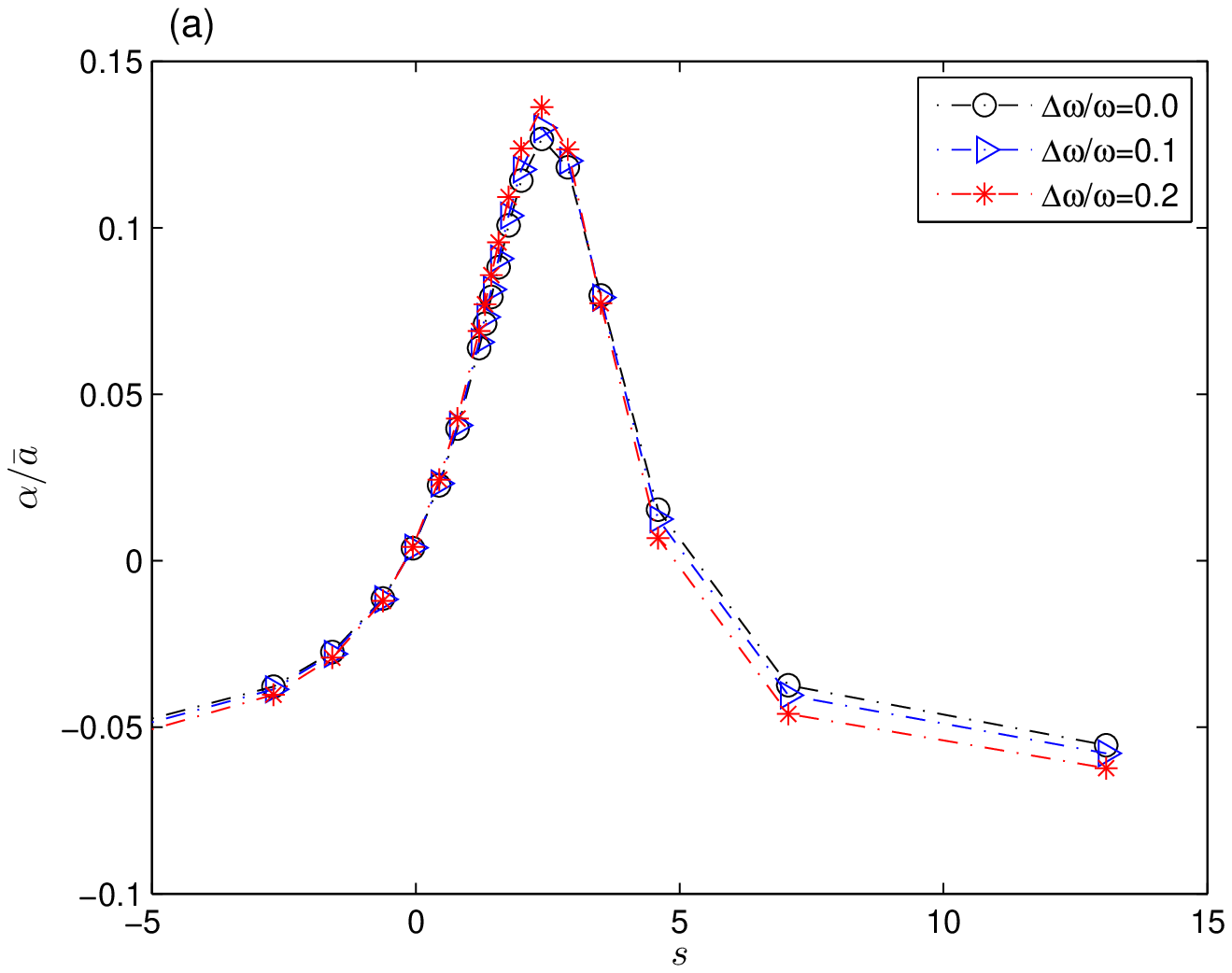}}
\end{subfigure}\\
\begin{subfigure}{
\includegraphics[width=0.7\columnwidth]{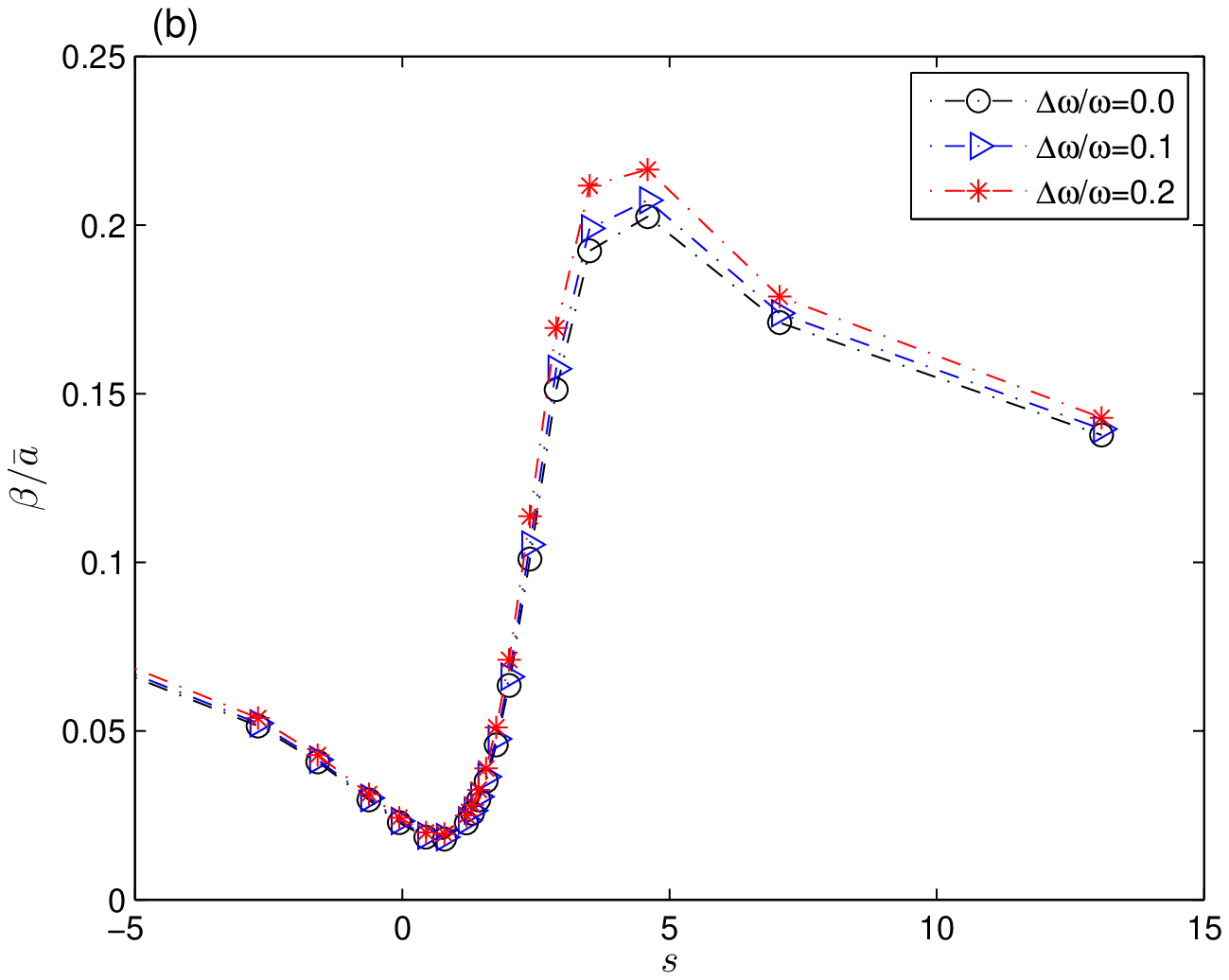}}
\end{subfigure}
\end{center}
\caption{(Color online) The effect of adding a slight anisotropy to the confining harmonic potential on (a)real and (b)imaginary part of the one dimensional scattering length as a function of 3D scattering length $s$ when $k_0 \rightarrow0$ for a loss parameter $y=0.3$. The frequency along the $y$ axis is fixed at $\omega_y=0.02 \omega_0$.}
\label{fig:Fig9}
\end{figure}

 In \figref{Fig7} we show the reaction rate constant graph at low energies just
 above the first channel threshold ($K^{1D,re}={\left( 1-{\left| S_{00} \right|}^2 \right)}/{(2k_0a_{\perp}^2)}$) for a special loss parameter $y=0.3$. Here, following the definition of the dimensionless reactive constant presented in \cite{Idziaszek2015}, we have divided $K^{1D,re}$ in Eq.(\ref{14}) by $2k_0^2a_{\perp}^2$. Our results show a good agreement with the low energy reaction constant formula
 which is suggested in \cite{Idziaszek2015}
\begin{eqnarray}\label{re_K}
K^{1D,re}=\frac{2+s(s-2)}{s^2+y^2{(s-2)}^2}
\end{eqnarray}
In \figref{Fig8} we present the one dimensional scattering length ($a_{1D}=\tilde {\alpha}+i\tilde {\beta}$) which we
have obtained numerically
for a slight anisotropy $\eta=\omega_x/\omega_y=1.05$ and $y=0.3$ along with the values predicted by the
following formula \cite{Peng}, for comparison
\begin{eqnarray}
a_{1D}=-\frac{a_{\perp}^2}{2\sqrt{\eta}\tilde{s}}\left( 1-C(\eta)\frac{\tilde{s}}{a_{\perp}} \right)
\end{eqnarray}
Here, $\tilde{s}$ is a complex number for the 3D scattering length in the presence of a loss mechanism and we have assumed $C(\eta=1.05)\approx C(\eta=1.0)=1.4603$. Our values show a good agreement with the analytical predictions.
We show the effect of adding a slight anisotropy on one dimensional scattering length in \figref{Fig9}. The energy is assumed to be very small, just above the first channel energy threshold. As it is observed, when the anisotropy is raised a little bit, it does not have a great effect on the real part of the scattering length ($\alpha$) and it can only result in a small increase of the imaginary part ($\beta$) mainly around the CIR position ($a_{\perp}/s=1.46$).
\begin{figure}[H]
\begin{center}
\begin{subfigure}{
\includegraphics[width=0.7\columnwidth]{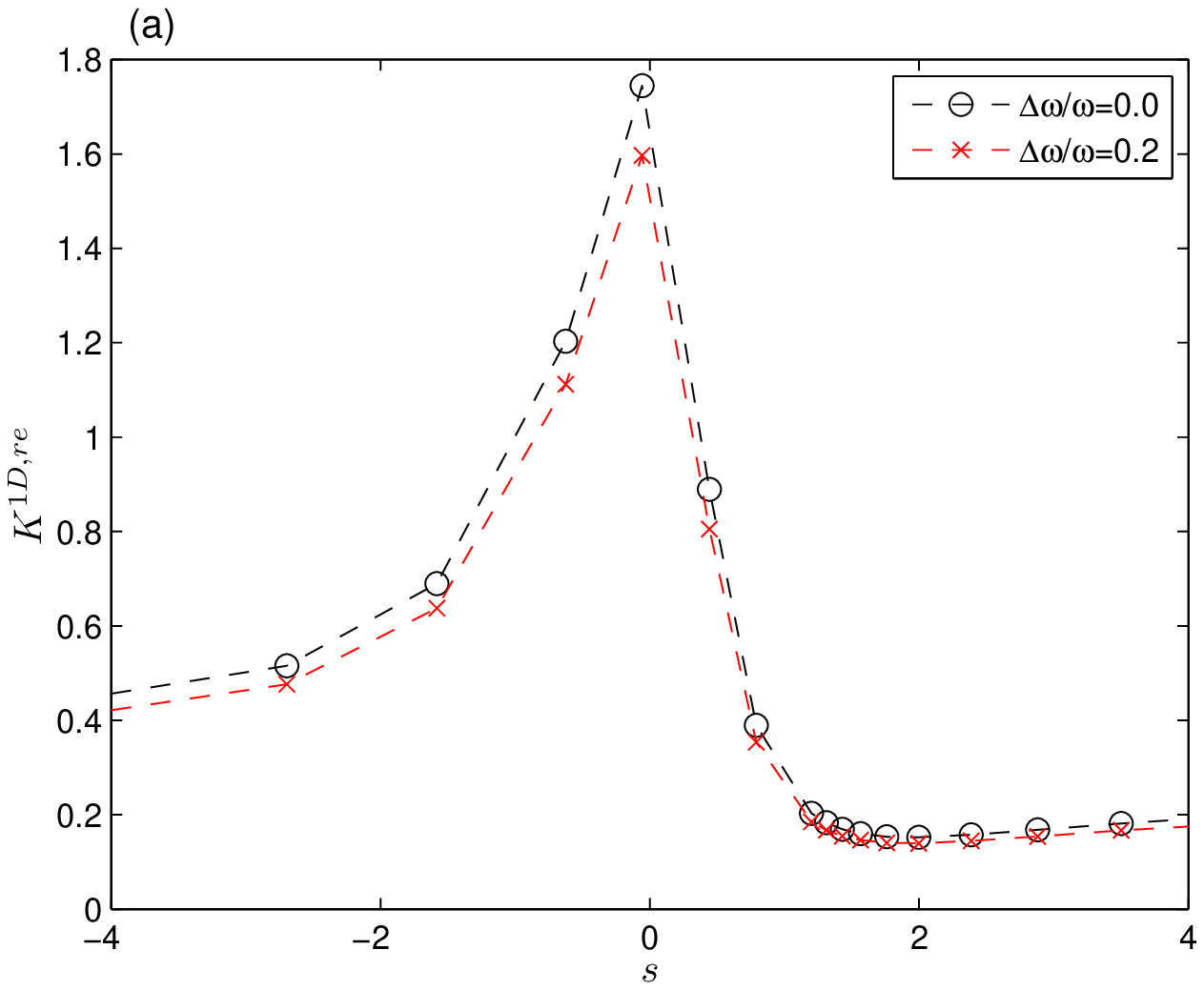}}
\end{subfigure}\\
\begin{subfigure}{
\includegraphics[width=0.7\columnwidth]{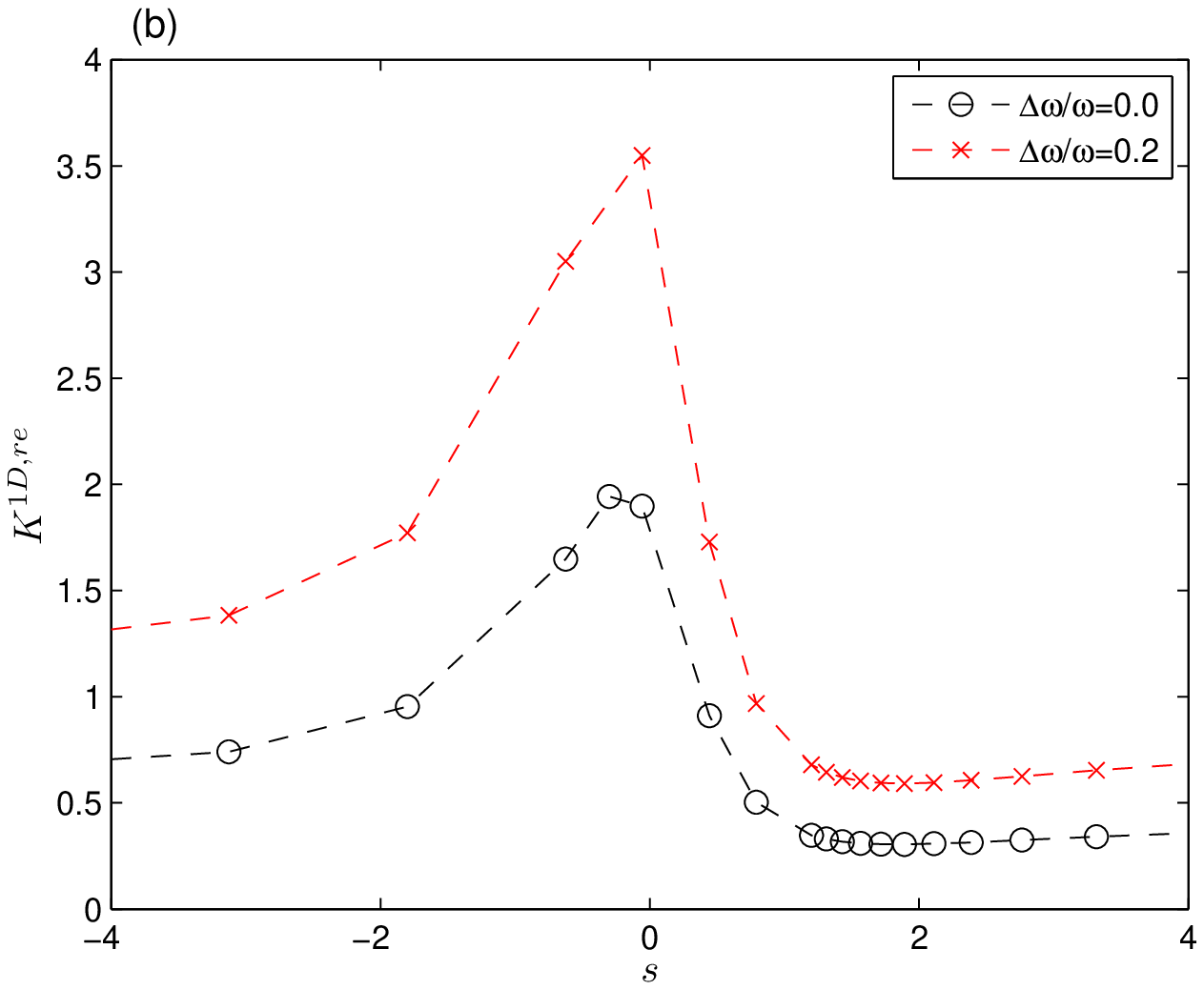}}
\end{subfigure}
\end{center}
\caption{(Color online) Reactive rate constant for anisotropic and isotropic harmonic waveguide for the cases of (a) one open channel (b) two open channels as a function of 3D scattering length $s$ when the energy is just above the last open channel threshold and the initial state is in the highest open channel ($k_0 \rightarrow0$ in (a) and $k_1 \rightarrow0$ in (b)) for a loss parameter $y=0.3$. The frequency along the $y$ axis is fixed at $\omega_y=0.02\omega_0$.}
\label{fig:Fig10}
\end{figure}
The anisotropy effect on the reactive rate when there is only one open channel differs from the case of two open channels.
We illustrate this phenomenon in \figref{Fig10}, when the initial energy of the system is slightly above the last open channel threshold. When only one transversal harmonic state is involved, there is no considerable impact on $K^{re}$ (\figref{Fig10} (a)). But for the multichannel scattering, as it can be seen in \figref{Fig10} (b), anisotropy yields a considerable increase in the $K^{re}$ value. However, the maximum appears around $s=0$ for both cases.
\section{Conclusions}
We have developed the computational scheme for ultracold
multichannel processes in harmonic waveguides with
transverse anisotropy. It is applicable to the case of elastic as
well as inelastic (reactive) multichannel scattering. By considering the effect of the transmission between higher transversal harmonic states, we have shown that a CIR splitting can occur for anisotropic harmonic confinement \cite{Melezhik2011} only if the interaction potential between colliding particles has long-range character with respect to the width of the confining trap. For the case of inelastic scattering, we have
reproduced the results of \cite{Idziaszek2010, Idziaszek2015} by
assuming a proper absorbing short range potential and studied the
multichannel issue in the presence of anisotropy, as well. We have
shown that although a slight anisotropy does not have a great
impact on inelastic one channel scattering, it can cause a
considerable change in the reactive rate constant when the
contribution of higher open channels is also included.
\section{Acknowledgements}
S.Sh would like to acknowlege the financial support by the Ministry of Science, Research and Technology of Iran. S.Sh thanks the Bogoliubov Laboratory of Theoretical Physics of JINR for their financial support and warm hospitality.

\section{Appendix A}
In Eq.(\ref{25}), each coefficient $A_{j^\prime}^j$ is a $N\times N$ matrix, each $\alpha_{j}$ ia a diagonal $N\times N$ matrix, and each $\gamma_{j}$ is a $N$-dimensional vector. \\
For up to three open channels $\left(n_{x},n_{y}\right)=\left(0,0\right),\left(0,2\right)$ and $\left(2,0\right)$, we have

\begin{eqnarray}
\alpha_{j}^{\left(1\right)}&=&-\frac{r_{j}}{r_{j-1}}\frac{\phi_{n_{e}}\left(x_{j},y_{j}\right)}{\phi_{n_{e}}\left(x_{j-1},y_{j-1}\right)}e^{ik_{n_{e}}\left(\left|z_{j}\right|-\left|z_{j-1}\right|\right)}\left(1+W_{0}^{\left(j\right)}+W_{1}^{\left(j\right)}\right),\nn\\
\alpha_{j}^{\left(2\right)}&=&\frac{r_{j}}{r_{j-2}}\frac{\phi_{n_{e}}\left(x_{j},y_{j}\right)}{\phi_{n_{e}}\left(x_{j-2},y_{j-2}\right)}e^{ik_{n_{e}}\left(\left|z_{j}\right|-\left|z_{j-2}\right|\right)}\left(W_{0}^{\left(j\right)}+W_{1}^{\left(j\right)}+W_{1}^{\left(j\right)}W_{0}^{\left(j-1\right)}\right),\nn\\
\alpha_{j}^{\left(3\right)}&=&-\frac{r_{j}}{r_{j-3}}\frac{\phi_{n_{e}}\left(x_{j},y_{j}\right)}{\phi_{n_{e}}\left(x_{j-3},y_{j-3}\right)}e^{ik_{n_{e}}\left(\left|z_{j}\right|-\left|z_{j-3}\right|\right)}W_{1}^{\left(j\right)}W_{0}^{\left(j-1\right)}
\end{eqnarray}

and

\begin{eqnarray}
\gamma_{j}&=&\sqrt{\lambda_{j}}r_j\phi_{n_e}\left(x_{j},y_{j}\right)e^{ik_{n_e}\left| z_j \right|} \left\{e^{-ik_{n_e}\left| z_j \right|}\cos(k_qz_j)\frac{\phi_q(x_j,y_j)}{\phi_{n_e}(x_j,y_j)} \right. \nn\\
&-&e^{-ik_{n_e}\left| z_{j-1} \right|}\cos(k_qz_{j-1})\frac{\phi_q(x_{j-1},y_{j-1})}{\phi_{n_e}(x_{j-1},y_{j-1})}(1+W_{0}^{(j)}+W_1^{(j)})\nn\\
&+&e^{-ik_{n_e}\left| z_{j-2} \right|}\cos(k_qz_{j-2})\frac{\phi_q(x_{j-2},y_{j-2})}{\phi_{n_e}(x_{j-2},y_{j-2})}(W_{0}^{(j)}+W_1^{(j)}+W_1^{(j)}W_0^{(j-1)})\nn\\
&-&\left. e^{-ik_{n_e}\left| z_{j-3} \right|}\cos(k_qz_{j-3})\frac{\phi_q(x_{j-3},y_{j-3})}{\phi_{n_e}(x_{j-3},y_{j-3})}W_1^{(j)}W_0^{(j-1)}\right\}
\end{eqnarray}

where $q$ and $n_e$ refer to the initial transverse state and the highest open channel,  respectively.\\
We define the following relations
\begin{eqnarray}
\chi_j&=&e^{i(k_{n_e-1}-k_{n_e})\left| z_j \right|}\frac{\phi_{n_e-1}(x_j,y_j)}{\phi_{n_e}(x_j,y_j)}-e^{i(k_{n_e-1}-k_{n_e})\left| z_{j-1} \right|}\frac{\phi_{n_e-1}(x_{j-1},y_{j-1})}{\phi_{n_e}(x_{j-1},y_{j-1})}\nn\\
\eta_j&=&e^{i(k_{n_e-2}-k_{n_e})\left| z_j \right|}\frac{\phi_{n_e-2}(x_j,y_j)}{\phi_{n_e}(x_j,y_j)}\nn\\
&-&e^{i(k_{n_e-2}-k_{n_e})\left| z_{j-1} \right|}\frac{\phi_{n_e-2}(x_{j-1},y_{j-1})}{\phi_{n_e}(x_{j-1},y_{j-1})}(1+\frac{\chi_j}{\chi_{j-1}})\nn\\
&+&\frac{\chi_j}{\chi_{j-1}}e^{i(k_{n_e-2}-k_{n_e})\left| z_{j-2} \right|}\frac{\phi_{n_e-2}(x_{j-2},y_{j-2})}{\phi_{n_e}(x_{j-2},y_{j-2})}
\end{eqnarray}


In the case of three open channels, the highest open channel is $(n_{e_x},n_{e_y})=(2,0)$ and we have

\begin{eqnarray}
W_0^{(j)}&=&\frac{\chi_j}{\chi_{j-1}}\nn\\W_1^{(j)}&=&\frac{\eta_j}{\eta_{j-1}},
\end{eqnarray}

while for two open channels, $(n_{e_x},n_{e_y})=(0,2)$ and

\begin{eqnarray}
W_0^{(j)}&=&0\nn\\W_1^{(j)}&=&\frac{\chi_j}{\chi_{j-1}}
\end{eqnarray}

finally single channel regime, $(n_{e_x},n_{e_y})=(0,0)$

\begin{eqnarray}
W_0^{(j)}&=&0\nn\\W_1^{(j)}&=&0
\end{eqnarray}

\end{document}